\documentclass[a4paper,11pt]{article}
\usepackage{pos}

\title{Estimates for the lightest baryon masses in $\mathcal{N} = 1$ supersymmetric Yang-Mills theory}

\usepackage{subfigure}
\usepackage{braket}
\newcommand{\arxiv}[2]{[arXiv:\,\href{http://arxiv.org/abs/#1}
{\texttt{#1}}[\texttt{#2}]]}
\newcommand{\arxivold}[1]{[arXiv:\,\href{http://arxiv.org/abs/#1}
{\texttt{#1}}\,]}
\newcommand{\I}{\ensuremath{\mathrm{i}}}

\newcommand{\mg}{m_{\tilde{g}}}

\author*[a,b]{Sajid Ali}%
\author[c]{Georg Bergner}
\author[c]{Camilo L\'opez}
\author[d]{Istvan Montvay}
\author[e]{Gernot M\"unster}
\author[f]{Stefano Piemonte}
\affiliation[a]{University of Bielefeld, Faculty of Physics,\\
Universit\"atsstr.~25, D-33615 Bielefeld, Germany}
\affiliation[b]{Government College University Lahore, Department of Physics,\\
Lahore 54000, Pakistan}
\affiliation[c]{University of Jena, Institute for Theoretical Physics, \\
Max-Wien-Platz 1, D-07743 Jena, Germany}
\affiliation[d]{Deutsches Elektronen-Synchrotron DESY, \\
Notkestr.~85, D-22607 Hamburg, Germany}
\affiliation[e]{University of M\"unster, Institute for Theoretical Physics,\\ 
Wilhelm-Klemm-Str.~9, D-48149 M\"unster, Germany}
\affiliation[f]{University of Regensburg, Institute for Theoretical Physics, \\
Universit\"atsstr.~31, D-93040 Regensburg, Germany}

\emailAdd{sajid.ali@physik.uni-bielefeld.de}
\emailAdd{georg.bergner@uni-jena.de}
\emailAdd{camilo.lopez@uni-jena.de}
\emailAdd{montvay@mail.desy.de}
\emailAdd{munsteg@uni-muenster.de}
\emailAdd{stefano.piemonte@ur.de}

\abstract{$\mathcal{N} = 1$ supersymmetric Yang-Mills theory describes
gluons interacting with gluinos, which are spin-$\frac{1}{2}$ Majorana
particles in the adjoint representation of the gauge group. In addition
to glueballs and mesonic bound states, the theory contains color
neutral bound states of three gluinos, which are analogous to baryons
in QCD. We calculate their correlation functions, involving
``sunset diagrams'' and ``spectacle diagrams'', numerically for gauge
group SU(2) and present an update on the estimates for the lowest masses.}

\FullConference{%
 The 38th International Symposium on Lattice Field Theory, LATTICE2021
  26th-30th July, 2021
  Zoom/Gather@Massachusetts Institute of Technology
}

\begin{document}
\maketitle

\section{Introduction}

$\mathcal{N}=1$ supersymmetric Yang-Mills (SYM) theory describes the
strong interaction between gluons and their superpartners, the gluinos,
represented by the fields $A^a_\mu(x)$ and $\lambda^a(x)$, respectively,
where $a=1, \ldots, N^2_c-1$ for gauge group SU($N_c$). Gluinos are Majorana fermions transforming
in the adjoint representation of the gauge group. The Lagrangian of
$\mathcal{N}=1$ SYM theory in Minkowski space reads
\begin{equation}
\mathcal{L}_{\text{SYM}} = -\frac{1}{4} F^a_{\mu\nu} F^{a,\mu\nu} 
+ \frac{\I}{2} \bar{\lambda}^a \gamma^\mu \left( \mathcal{D}_\mu \lambda \right)^a 
- \frac{\mg}{2} \bar{\lambda}^a \lambda^a .
\end{equation}
The first term in the above Lagrangian is the gauge part where
$F^a_{\mu\nu}$ is the non-abelian field strength tensor. The second term
is the kinetic part, where $\mathcal{D}_\mu$ is the covariant
derivative in the adjoint representation. The last term is the gluino mass
term with mass $\mg$, which breaks supersymmetry softly. 
For the Monte-Carlo simulations on an Euclidean
hypercubic lattice we use the action proposed by Curci and Veneziano
\cite{Curci:1986sm} based on a Wilson fermion formulation. 
The bare mass is related the hopping parameter by
$\kappa = 1/(2 \mg + 8)$. 
A clover term can be added to the action to improve
the results up to $O(a^2)$~\cite{Musberg:2013foa}. The gluino mass has to be finite 
in the simulations, and a chiral extrapolation 
is required to obtain final results in zero gluino mass limit.

Based on effective Lagrangians, Veneziano et al.\ and Farrar et al.\ predicted
supermultiplets of lightest bound states for the theory \cite{Veneziano:1982ah,Farrar:1997fn}. 
The masses of the members of a multiplet are degenerate if supersymmetry is not broken.

In the last few years we have investigated the low-lying mass
spectrum of SYM theory on the lattice with gauge groups SU(2) and
SU(3), which we have calculated nonperturbatively from first principles
using Monte-Carlo techniques
\cite{Bergner:2015adz,Ali:2017iof,Ali:2018dnd,Ali:2019gzj,Ali:2019agk}.
In addition, we have studied the SUSY Ward identities, where we showed
that the broken supersymmetry is recovered in the combined chiral and continuum limit
\cite{Ali:2018fbq,Ali:2020mvj}.

Theoretically it is possible in SYM theory to form bound states of three gluinos,
for any number of colors and for gauge groups SU(2) and SU(3). In
analogy to the baryons of QCD, we also call these objects generally
``baryons''. We have already presented preliminary results for baryon
masses for the gauge group SU(2)~\cite{Ali:2019baryon}. In this article
we will present improved results, the improvement being made by choosing an
optimal set of parameters in the stochastic estimator technique and by putting
more sources on the full time extent of the lattice. In Sec.~\ref{corr}
we will briefly discuss the theoretical construction of baryon correlation
functions, while Sec.~\ref{num} will be about numerical results of
correlation functions and corresponding masses.

\section{Baryon correlation functions}\label{corr}

As a first step, we discuss the analytical form of the baryon two-point
correlation function that is numerically evaluated on the lattice and
the appropriate function fitted to the data to obtain the corresponding mass.
The correlation function can be obtained from the interpolating field
$W(x)$ and its conjugate field $\overline{W}(x)$ as follows
\begin{equation}
B(x,y) = \braket{W(x)\overline{W}(y)}.
\end{equation}
The interpolating field $W(x)$ is a three fermion operator with by now unspecified matrices
$\Gamma$ acting on spinor indices.
Using the Majorana condition, the correlator 
can be simplified to \cite{Ali:2019thesis}
\begin{align}
B^{\alpha \delta}(x,y)
& = \langle W^\alpha(x) \overline{W}^{\delta}(y) \rangle 
= \langle W^{\alpha}(x) C^{\delta \alpha'} W^{\alpha'}(y) \rangle \nonumber\\
& = t_{abc} t_{a'b'c'} \Gamma^{\beta\gamma} \Gamma^{\beta'\gamma'}
C^{\delta \alpha'}
\, \langle
\lambda_a^{\alpha}(x)\lambda_b^{\beta}(x)
\lambda_c^{\gamma}(x)\lambda_{a'}^{\alpha'}(y)\lambda_{b'}^{\beta'}(y)
\lambda_{c'}^{\gamma'}(y)\rangle,
\end{align}
where $C$ is the charge conjugation matrix.
The fermions are integrated out based on Wick's theorem. 
For the contractions we use
$\langle \lambda^{\alpha}_a(x) \lambda^{\beta}_b(y) \rangle_F
= -\left(\Delta(x,y) C \right)^{\alpha\beta}_{ab}$,
where the propagator $\Delta = D_w^{-1}$ is the inverse of the Wilson-Dirac matrix $D_w$. 
We obtain two major contributions,
namely the ``sunset piece'' ($B^{\pm}_{Sset}$) and the ``spectacle piece'' ($B^{\pm}_{Spec}$). 
\begin{figure}[hbt!]
\centering
\includegraphics[width=0.35\textwidth]{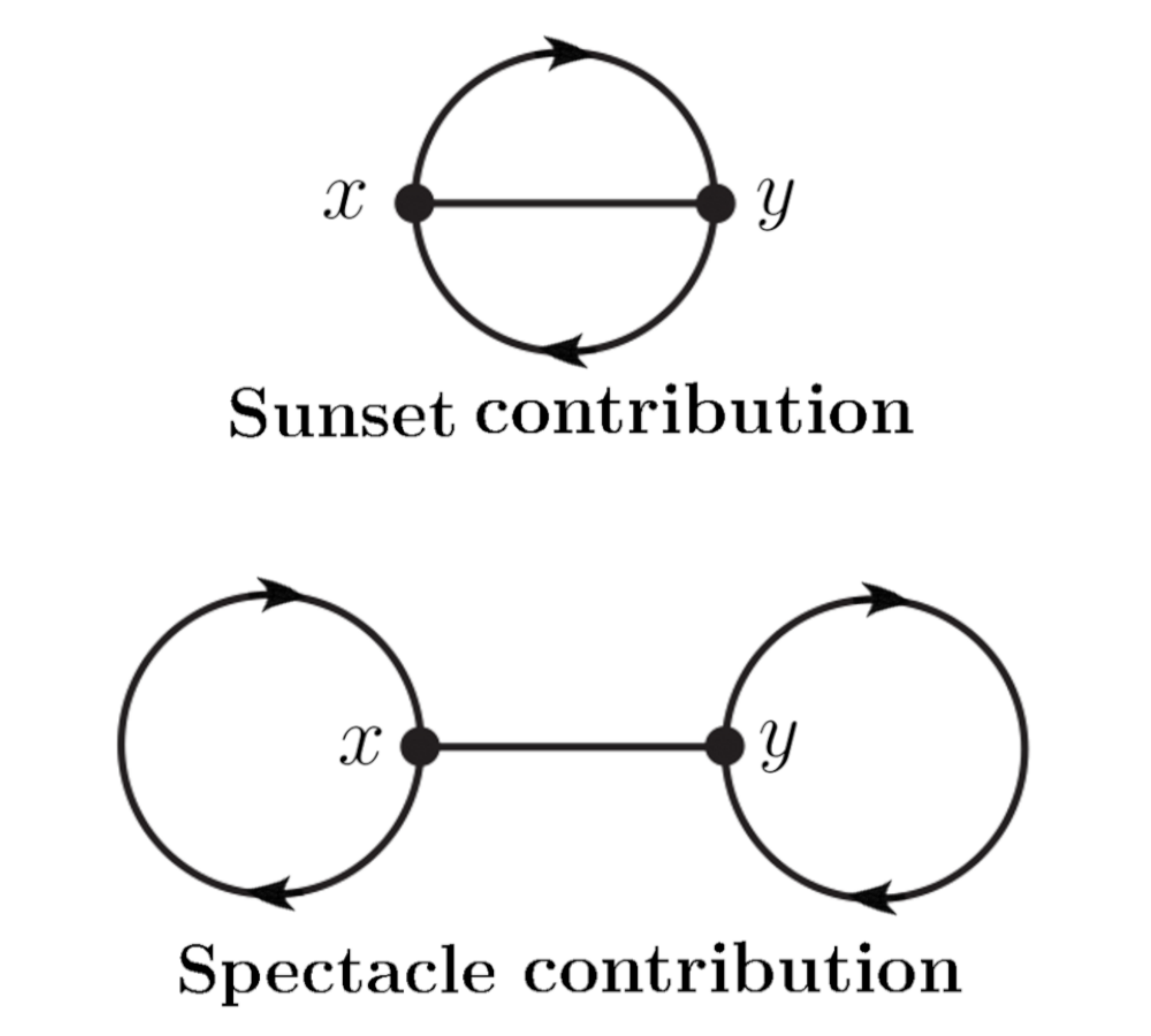}
\caption{Graphical representation of the sunset and spectacle contributions.}
\end{figure}
With the choice $\Gamma = C\gamma_4$, and for gauge group SU(2), 
which we consider in the numerical work,
the expressions are
\begin{align}
B^{\pm}_{Sset}(x,y) = -\varepsilon_{abc} \varepsilon_{a'b'c'} &(C\gamma_4)^{\beta\gamma} (C\gamma_4)^{\beta'\gamma'} P_{\pm}^{\alpha\alpha'}
\nonumber\\
\langle &  + 2 \Delta^{\alpha \alpha'}_{a a'}(x,y)\Delta^{\beta \beta'}_{b b'}(x,y)\Delta^{\gamma \gamma'}_{c c'}(x,y)\nonumber\\
&  + 4 \Delta^{\alpha \beta'}_{a b'}(x,y) \Delta^{\beta \gamma'}_{b c'}(x,y) \Delta^{\gamma \alpha'}_{ca'}(x,y)
\rangle,
\end{align}
and
\begin{align}\label{Spec}
B^{\pm}_{Spec}(x,y) = -\varepsilon_{abc} \varepsilon_{a'b'c'} &(C\gamma_4)^{\beta\gamma} (C\gamma_4)^{\beta'\gamma'} P_{\pm}^{\alpha\alpha'} 
\nonumber\\
\langle &  + 2 \Delta^{\alpha \beta}_{ab}(x,x)   \Delta^{\delta \alpha'}_{c a'}(x,y)\Delta^{\delta' \beta'}_{c' b'}(y,y)C^{\gamma \delta}C^{\delta' \gamma'}\nonumber\\
&  + 4 \Delta^{\alpha \beta}_{a b}(x,x)   \Delta^{\beta' \gamma}_{b' c}(y,x) \Delta^{\gamma' \alpha'}_{c' a'}(y,y)\nonumber\\
&  + 1 \Delta^{\alpha \alpha'}_{a a'}(x,y)\Delta^{\beta \delta}_{b c}(x,x)   \Delta^{\delta' \beta'}_{c' b'}(y,y) C^{\gamma\delta} C^{\delta' \gamma'}\nonumber\\
&  + 2 \Delta^{\alpha \delta'}_{a c'}(x,y) \Delta^{\beta \delta}_{b c}(x,x)   \Delta^{\beta' \alpha'}_{b' a'}(y,y) C^{\gamma\delta} C^{\delta' \gamma'}
\rangle.
\end{align}
Here $P_{\pm}$ denotes a parity projection; 
for zero momentum states it is defined as
$P_{\pm} = \frac{1}{2} (1 \pm \gamma_4)$ \cite{MontvayMuenster}.
$\varepsilon_{abc}$ are the totally antisymmetric structure constants for SU(2). 

\section{Numerical results}\label{num}

\begin{figure}[hbt!]
\centering
\subfigure[Sunset contribution only.]{\label{FigTotCorr}
\includegraphics[width=0.47\textwidth]{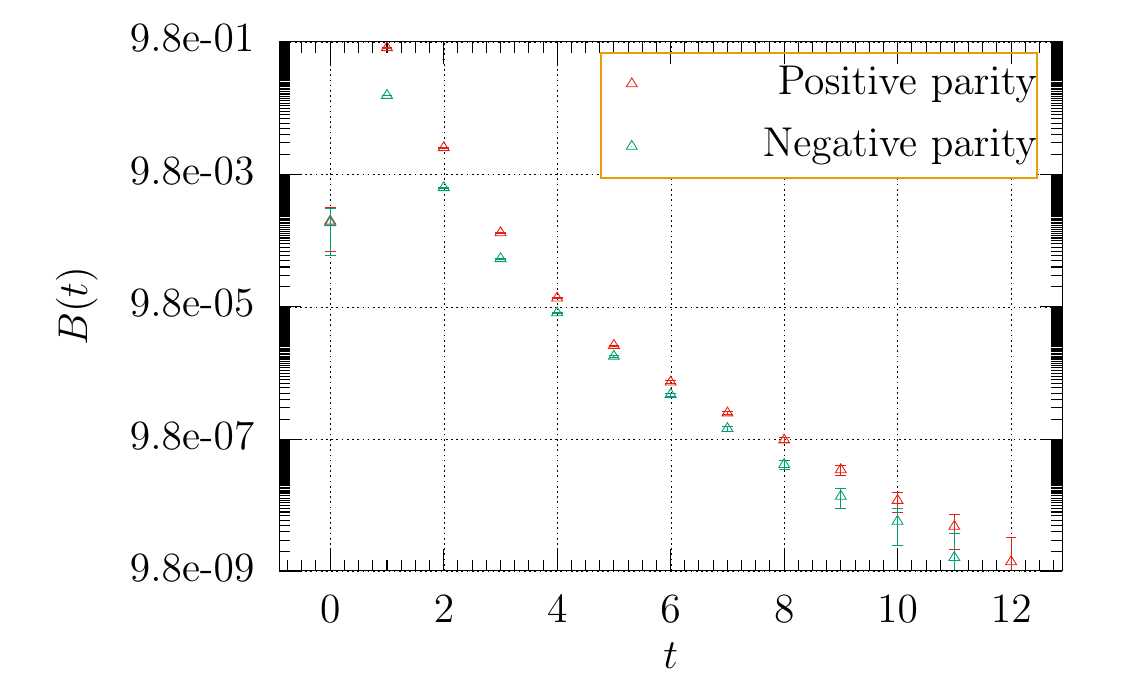}}\quad
\subfigure[Both sunset and spectacle contributions.]{\label{FigTotEmass}
\includegraphics[width=0.47\textwidth]{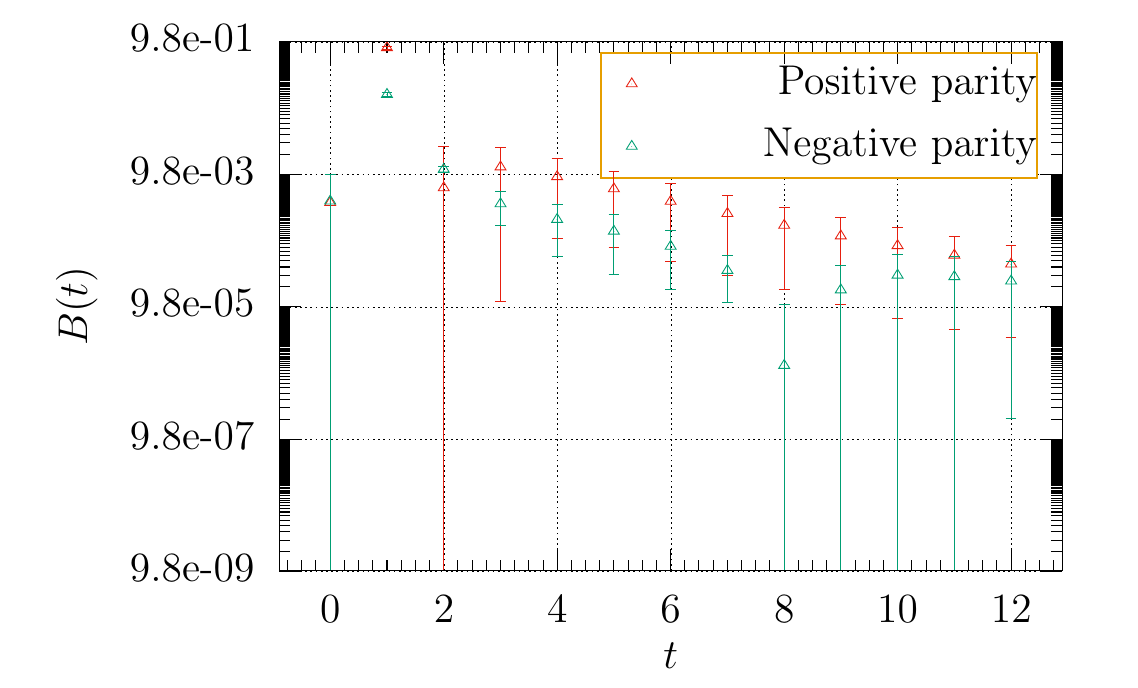}}
\caption{Numerical results for the positive and negative parity baryon
correlation functions at $\beta=1.75$ and $\kappa=0.14925$.}
\label{PparNparCorr}
\end{figure}

\begin{figure}[hbt!]
\centering
\subfigure[Sunset: Effective mass of positive parity baryon.]{\label{FigTotCorr}
\includegraphics[width=0.47\textwidth]{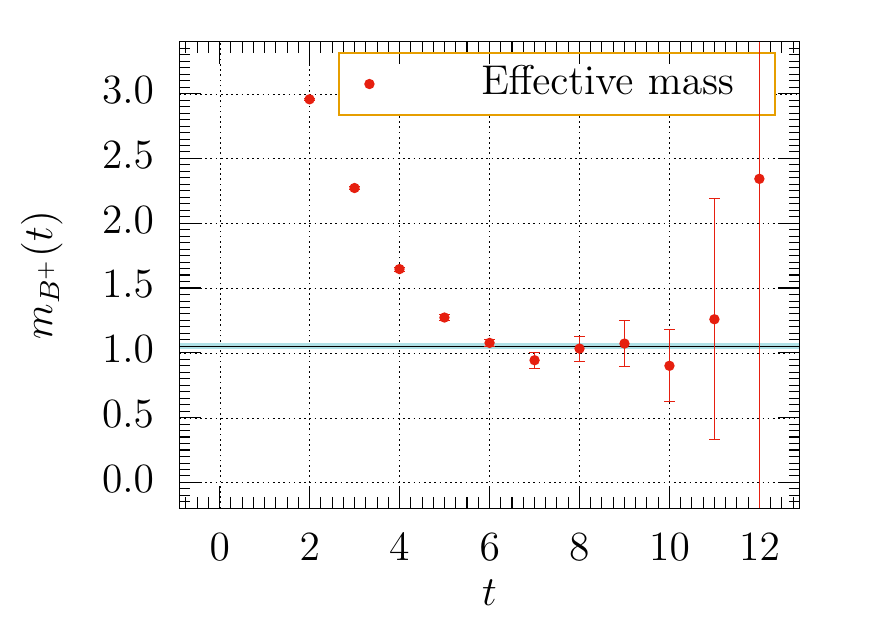}}\quad
\subfigure[Sunset: Effective mass of negative parity baryon.]{\label{FigTotEmass}
\includegraphics[width=0.47\textwidth]{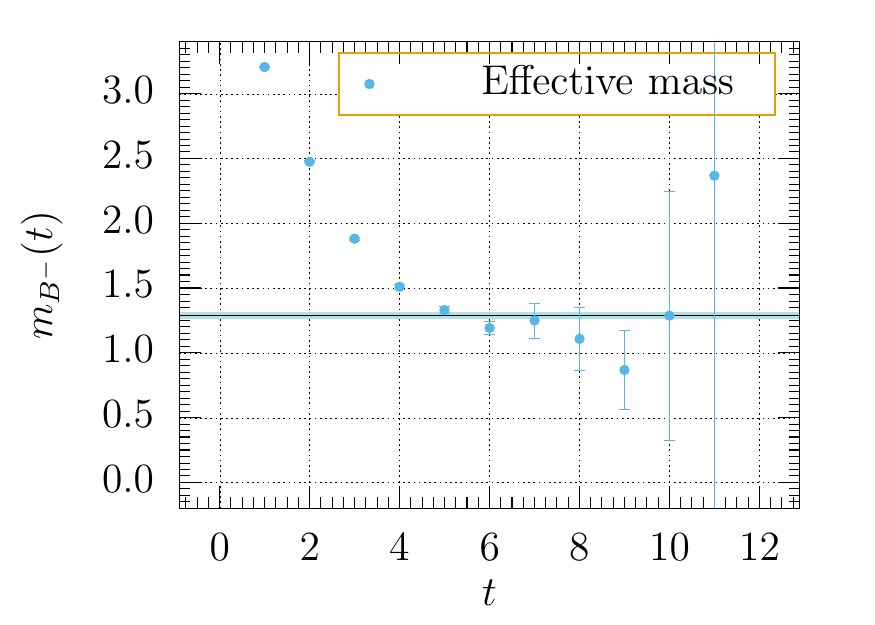}}\quad
\subfigure[Full: Effective mass of positive parity baryon.]{\label{FigTotEmass}
\includegraphics[width=0.47\textwidth]{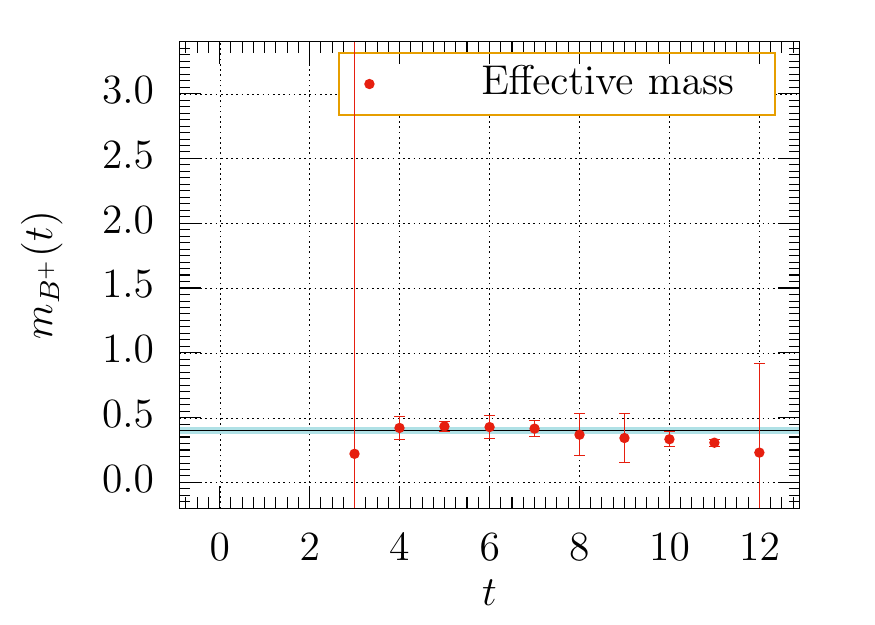}}\quad
\subfigure[Full: Effective mass of negative parity baryon.]{\label{FigTotEmass}
\includegraphics[width=0.47\textwidth]{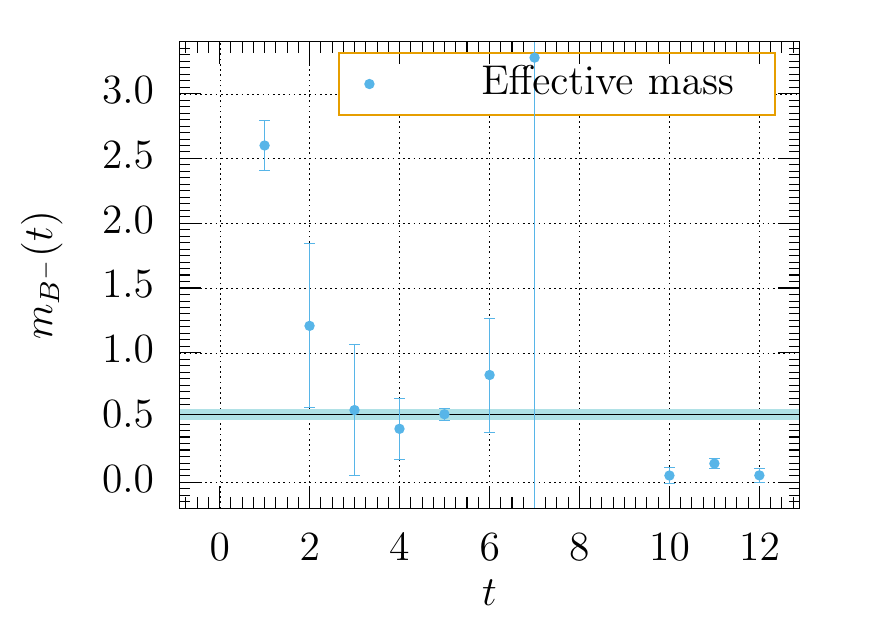}}
\caption{Numerical results for the effective masses at $\beta=1.75$ and
$\kappa=0.14925$.}
\label{emass}
\end{figure}

In the present section numerical results for baryon correlation functions
and corresponding masses for gauge group SU(2) will be discussed. 
The numerically intensive and
non-trivial task is to compute the inverse of the Wilson-Dirac operator. 
The fermion propagators appearing in the
``sunset piece'' are obtained from point sources using standard techniques. 
The all-to-all propagators appearing
in the ``spectacle piece'' are calculated using the stochastic estimator
technique. A combination of 80 stochastic estimators and
200 lowest eigenvalues has been used to minimize the stochastic noise.
These parameters are optimized compared to the first test presented in
Ref.~\cite{Ali:2019baryon} to improve the signal. As the spectacle
contribution is still noisy, we used several sources placed randomly at all
time slices of the lattice. After computing both contributions separately, 
they are added up for each configuration to
obtain the full correlation function. We measured the baryon correlator for
lattice volume $24^3\times48$ at $\beta=1.75$ and $\kappa=0.14925$ on 1757
configurations. Numerical results for positive and negative parity correlation
functions are shown in Fig.~\ref{PparNparCorr}. Statistical errors have been
estimated by the Jackknife method.
To remove the autocorrelation along the Monte-Carlo time
the data blocking method, with an optimal binlength of 50,
has been used.
In order to obtain estimates for the masses of baryonic states, the
effective mass of positive and negative parity baryons is plotted as a
function of time-slice distance, and is shown in Fig.~\ref{emass}. For precise
results, a $\sinh$ function is fitted to the numerical data and a fit range
$t\in \{4,9\}$ for both parities is considered. Masses obtained from the above
fit ranges are shown in Tab.~\ref{massesTab}.
\begin{table}[hbt!]
\begin{center}
\begin{tabular}{|c||c|}
\hline
Bound state  &  Mass (in lattice units)\\
\hline
\hline
$m_{B^{+}}$  &  0.4207(083) \\
\hline
$m_{B^{-}}$  &  0.5429(263)\\
\hline
\end{tabular}
\caption{Masses of the positive and negative parity baryons in
$\mathcal{N}$=1 SUSY Yang-Mills theory at $\beta=1.75$ and $\kappa=0.14925$.}
\label{massesTab}
\end{center}
\end{table}
%
\section{Conclusion and outlook}

We presented numerical results for positive and
negative parity baryon correlators and their corresponding masses in lattice
units for the gauge group SU(2). Both masses are different as expected.
They are still at finite gluino mass and lattice spacing. In the
next step we will measure the baryon masses from ensembles produced at
different bare gluino masses, which allows to perform chiral extrapolations.
In the future we plan to measure the masses for the gauge group SU(3),
where we have better statistics and a well controlled continuum limit.

\section*{Acknowledgements}

The authors gratefully acknowledge the Gauss Centre for Supercomputing
e.~V.\ (www.gauss-centre.eu) for funding this project by providing computing
time on the GCS Supercomputer JUQUEEN and JURECA at J\"ulich Supercomputing
Centre (JSC) and SuperMUC at Leibniz Supercomputing Centre (LRZ). Further
computing time has been provided on the compute cluster PALMA of the
University of M\"unster. This work is supported by the Deutsche
Forschungsgemeinschaft (DFG) through the Research Training Group ``GRK 2149:
Strong and Weak Interactions -- from Hadrons to Dark Matter''. G.~Bergner\
and C.~Lopez acknowledge support from the Deutsche Forschungsgemeinschaft (DFG)
under Grant No.\ BE 5942/2-1 and BE 5942/3-1. S.~Ali acknowledges financial support from the
Deutsche Akademische Austauschdienst (DAAD).


\end{document}